\documentclass[english,aps,prd,floats,floatfix,nofootinbib,preprintnumbers]{revtex4}
\usepackage{natbib}
\usepackage{times}
\usepackage{float}
\usepackage{amsmath}
\usepackage{graphicx}
\usepackage{amssymb}
\usepackage{slashed}
\usepackage[usenames]{xcolor}
\usepackage{soul} 
\usepackage{url}
\usepackage[colorlinks, hyperindex]{hyperref}
	\hypersetup
	{
		colorlinks, %
		citecolor=blue, %
		linkcolor=red, %
		urlcolor=blue, %
	}
\usepackage{doi}

\begin{document}
\preprint{CERN-TH-2017-085}
\title{Towards the discovery of new physics with lepton-universality ratios of $b\to s\ell\ell$ decays}
\author{Li-Sheng Geng$^1$, Benjam\'in Grinstein$^2$, Sebastian J\"ager$^3$, Jorge Martin Camalich$^4$, Xiu-Lei Ren$^{5,6}$ and Rui-Xiang Shi$^1$}
\affiliation{
$^1$School of Physics and Nuclear Energy Engineering and International Research Center for Nuclei and Particles in the Cosmos, Beihang University, Beijing 100191, China\\
$^2$Dept. Physics, University of California, San Diego, 9500 Gilman Drive, La Jolla, CA 92093-0319, USA\\
$^3$University of Sussex, Department of Physics and Astronomy, Falmer, Brighton BN1 9QH, UK\\
$^4$CERN, Theoretical Physics Department, CH-1211 Geneva 23, Switzerland\\
$^5$State Key Laboratory of Nuclear Physics and Technology, School of Physics, Peking University, Beijing 100871, China\\
$^6$Institut f\"ur Theoretische Physik II, Ruhr-Universit\"at Bochum, D-44780 Bochum, Germany}

\begin{abstract}
Tests of lepton-universality as rate ratios in $b\to s \ell\ell$ transitions can be predicted very accurately in the Standard Model. The deficits with respect to expectations reported by the LHCb experiment in muon-to-electron ratios of the $B\to K^{(*)}\ell\ell$ decay rates thus point to genuine manifestations of lepton non-universal new physics. In this paper, we analyse these measurements in the context of effective field theory. First, we discuss the interplay of the different operators in $R_K$ and $R_{K^*}$ and provide predictions for $R_{K^*}$ in the Standard Model and in new-physics scenarios that can explain $R_K$. We also provide approximate numerical formulas for these observables in bins of interest as functions of the relevant Wilson coefficients.
Secondly, we perform frequentist fits to $R_{K}$ and $R_{K^*}$.
The SM disagrees with these measurements at $3.7\sigma$ significance.
We find excellent fits in scenarios with combinations of $\mathcal O_{9(10)}^\ell=\bar s\gamma^\mu b_L~\ell\gamma_\mu(\gamma_5) \ell$ operators, with pulls relative  to the Standard Model in the region of $4\sigma$.
  An important conclusion of our analysis is that a lepton-specific contribution to $\mathcal O_{10}$ is important to understand the data. Under the hypothesis that new-physics couples selectively to the muons, we also present fits to other $b\to s\mu\mu$ data with a conservative error assessment, and comment on
    more general scenarios. Finally, we discuss new lepton universality ratios
  that, if new physics is the origin of the observed discrepancy,
    should contribute to the statistically significant discovery
    of new physics in the near future.        
\end{abstract}

\maketitle

\section{Introduction}

Ratios of decay rates such as $B\to K^{(*)}\ell\ell$ for different leptons $\ell=e$ or $\mu$ are protected from hadronic uncertainties and can be very accurately predicted in the Standard Model (SM)~\cite{Hiller:2003js,Bordone:2016gaq}. Therefore, significant discrepancies with experiment in these observables would have to be interpreted as unambiguous signals of new physics (NP) that, in addition, must be related to new lepton non-universal interactions. After first measurements reporting no significant deviation from universality (within large experimental uncertainties) at $B$-factories~\cite{Wei:2009zv,Lees:2012tva,Lees:2013nxa}, in 2014 the LHCb collaboration reported a quite precise measurement of the ratio $R_K=\Gamma(B\to K\mu\mu)/\Gamma(B\to Kee)$~\cite{Aaij:2014ora}
\begin{align}
R_K=0.745^{+0.090}_{-0.074}\pm0.036,\label{eq:RKLHCb}
\end{align}
in the bin corresponding to dilepton mass squared $q^2\in[1,~6]$ GeV$^2$ that deviated from the SM predictions $R_K=1.0004(8)$~\cite{Hiller:2003js,Bordone:2016gaq} with a significance of $2.6\sigma$. Recently, the LHCb reported the measurement of $R_{K^*}=\Gamma(B\to K^*\mu\mu)/\Gamma(B\to K^*ee)$ in two bins of $q^2$~\cite{Bifani}, 
\begin{align}
R_{{K^*}[0.045,~1.1]~\text{GeV}^2}=0.660^{+0.110}_{-0.070}\pm0.024,~~~~~R_{{K^*}[1.1,~6]~\text{GeV}^2}=0.685^{+0.113}_{-0.069}\pm0.047,\label{eq:RKsLHCb}
\end{align}
where the first error is statistical the second systematic, and which show again deficits with respect to the SM predictions (see below),
\begin{align}
R_{{K^*}[0.045,~1.1]~\text{GeV}^2}^{\rm SM}=0.920(7),~~~~~~~R_{{K^*}[1.1,~6]~\text{GeV}^2}^{\rm SM}=0.996(2),~\label{eq:RKsOurs} 
\end{align}
with a significance of $2.3\sigma$ and $2.4\sigma$, respectively.    

In addition, during the last few years there have been considerable investigations related to tensions with SM predictions in the angular analysis of $B\to K^*\mu\mu$~\cite{Aaij:2013qta,Aaij:2015oid,Wehle:2016yoi,ATLAS:2017dlm,CMSDinardo} and in branching fractions~\cite{Aaij:2014pli,Aaij:2015xza,Aaij:2015esa} whose significance have been claimed to be in the $4-5\sigma$ range in some global analyses~\cite{Beaujean:2013soa,Descotes-Genon:2015uva,Altmannshofer:2017fio}, although these observables are afflicted by hadronic uncertainties that obscure the interpretation and significance of the anomaly~\cite{Beneke:2001at,Grinstein:2004vb,Egede:2008uy,Khodjamirian:2010vf,Beylich:2011aq,Khodjamirian:2012rm,Jager:2012uw,Horgan:2013pva,Lyon:2014hpa,Descotes-Genon:2014uoa,Jager:2014rwa,Straub:2015ica,Ciuchini:2015qxb,Brass:2016efg,Aaij:2016cbx,Hurth:2016fbr,Capdevila:2017ert,Chobanova:2017ghn}. On the other hand, if the NP coupled selectively to the muons, then the effect that would be needed to explain the $R_K$ anomaly could also naturally explain these other anomalies in the angular distributions and rates~\cite{Alonso:2014csa,Hiller:2014yaa,Ghosh:2014awa,Hurth:2014vma}. Several authors have proposed combinations of the angular observables of $B\to K^{(*)}\ell\ell$ for muons and electrons probing lepton universality, that are again protected from hadronic uncertainties and sensitive to specific scenarios of NP~\cite{Altmannshofer:2014rta,Jager:2014rwa,Capdevila:2016ivx,Serra:2016ivr}. Indeed, some of these observables have been recently analysed by the Belle~\cite{Wehle:2016yoi} collaboration showing overall consistency with the SM, albeit with large experimental errors. 

In this paper, we perform an analysis of the $R_{K^*}$ measurements together with $R_K$ and other $b\to s\mu\mu$ data. First, we describe the dependence of the relevant lepton-universality ratios on lepton-specific Wilson coefficients of the weak Hamiltonian of $b\to s\ell\ell$ transitions, focusing in the correlations induced between $R_K$ and $R_{K^*}$. We then present numerical predictions for $R_K$ and $R_{K^*}$ in the SM and in benchmarks scenarios of NP for the bins of interest and present numerical formulas for their dependence on the Wilson coefficients. Finally, we present a series of fits to the data, including in subsequent steps the $B_s\to\mu\mu$ branching fraction and all the measurements of $CP$-averaged combinations of angular observables of $B\to K^*\mu\mu$ at low $q^2$. We describe our findings in terms of frequentist statistic inference and discuss the robustness of the results to variations of hadronic uncertainties.

\section{Structure of the new-physics contributions to $R_{K^{(*)}}$}

To leading order in $G_F$ the effective Hamiltonian for $b\to s\ell\ell$ transitions at low-energies ($\mu\sim m_b$) in the SM is~\cite{Grinstein:1988me,Buchalla:1995vs,Chetyrkin:1996vx}
\begin{equation}
\label{eq:Heff}
  {\cal{H}}_{\rm eff}^{\rm SM}=
  \frac{4 G_F}{\sqrt{2}}\sum_{p=u,c}\lambda_{ps}\left(C_1 \mathcal{O}_1^p+C_2\mathcal{O}_2^p +\sum_{i=3}^{10} C_i  \mathcal{O}_i\right),
\end{equation}
with $\lambda_{ps}=V_{pb}^{} V_{p s}^\ast$. The $\mathcal{O}_{7}$ and $\mathcal{O}_{9,10}$ are the electromagnetic penguin and the semileptonic operators, respectively, and $\mathcal{O}_{1,2}^p$,  $\mathcal{O}_{3,\ldots,6}$, and $\mathcal{O}_{8}$ are the ``current-current'', ``QCD-penguins'' and ``chromo-magnetic'' operators, respectively, which require of an electromagnetic interaction to contribute to the $b\to s\ell\ell$ transition  via ``non-factorizable'' corrections, in the language of QCD factorization~\cite{Beneke:2001at}. The effects of new physics beyond the SM can be modeled by modifying the Wilson coefficients $C_1,\ldots,C_{10}$ and by supplementing the effective Hamiltonian with chirally-flipped ($b_{L(R)}\to b_{R(L)}$) versions of these operators $\mathcal{O}_{7,\ldots, 10}'$, and also four scalar and two tensor operators~\cite{Bobeth:2007dw}.~\footnote{In the bottom-up approach followed in this paper one defines different ``new-physics scenarios'' by varying the values of these Wilson coefficients, without 
considering which could be the new degrees of freedom or specific UV completions of the SM producing them.}

Among all the possible operators present in $\mathcal H_{\rm eff}$, only the semileptonic ones,
\begin{eqnarray}
\mathcal{O}_9^{(\prime)\ell}=\frac{\alpha_{\rm em}}{4\pi}(\bar s\gamma^\mu P_{L(R)} b)\;(\bar\ell\gamma_\mu\ell),\hspace{1cm} 
\mathcal{O}_{10}^{(\prime)\ell}=\frac{\alpha_{\rm em}}{4\pi}(\bar s\gamma^\mu P_{L(R)} b)\;(\bar\ell\gamma_\mu\gamma_5\ell), \label{eq:semileptonic0}
\end{eqnarray}
can explain the deficit observed in $R_K$. The current-current, QCD penguin and magnetic operators do not induce lepton-universality violation because they connect to the dilepton pair through a photon. Scalar and tensor Lorentz structures cannot explain $R_K$ either if they stem from an UV completion of the SM manifesting at the electroweak scale as $SU(3)_c\times SU(2)_L\times U(1)_Y$-invariant effective operators: Tensor operators with the particle content of the SM of the type $(\bar Q_L\sigma_{\mu\nu}d_R)(\bar L_L\sigma^{\mu\nu} \ell_R)$ are forbidden by conservation of hypercharge, whereas $(\bar Q_L\sigma_{\mu\nu}d_R)(\bar \ell_R\sigma^{\mu\nu} L_L)$ is identically equal to zero. On the other hand, out of the four possible scalar operators at low energies only two are independent, and these are found to be severely constrained by the $B_q\to\ell\ell$ decay rates so they cannot explain $R_K$ either (see refs.~\cite{Alonso:2014csa,Hiller:2014yaa} for details). 

In terms of the operators of the type~(\ref{eq:semileptonic0}), and expanding around the massless limit for the lepton the differential decay rate of the $B\to K\ell\ell$ process is~\cite{Bobeth:2007dw}
\begin{align}
\frac{d\Gamma_K}{dq^2}=&\mathcal N_K|\vec k|^3 f_+(q^2)^2\left(\left|C_{10}^\ell+C_{10}^{\prime\ell}\right|^2+\left|C_9^\ell+C_9^{\prime\ell}+2\frac{m_b}{m_B+m_K}C_7\frac{f_T(q^2)}{f_+(q^2)}-8\pi^2 h_K\right|^2\right)\nonumber\\
&+\mathcal O(\frac{m_\ell^4}{q^4})+\frac{m_\ell^2}{m_B^2}\times O(\alpha_s,~\frac{q^2}{m_B^2}\times\frac{\Lambda}{m_b}),\label{eq:BtoKllrate}
\end{align}
where $\Lambda\equiv\Lambda_{\rm QCD}$, $\vec k$ is the 3-momentum of the recoiling meson in the $B$-meson rest-frame, $\mathcal N_K$ is a dimensionful normalization constant that drops out in in the ratio $R_K$, $f_{+,T}(q^2)$ are $B\to K$ form factors and $h_K$ encompasses the hadronic effects of the current-current, chromomagnetic and QCD penguin operators. It is clear from this equation that phase-space effects induced by the lepton masses are negligible as soon as $q^2\gtrsim 1$ GeV$^2$ and that the ratio of decay rates into muons and electrons must be very accurately equal to 1 (up to electromagnetic corrections~\cite{Bordone:2016gaq}), and regardless of hadronic contributions since these are necessarily lepton-universal. Since, $C_{1,\ldots,8}$ do not directly result in lepton-universality violation, any significant deficit from 1 in $R_K$ must be then caused by non-universal NP contributions $\delta C_{9,10}^{(')}$. Hence, taking into account that $C_9^{\rm SM}(m_b)\simeq-C^{\rm SM}_{10}=4.27$, explaining the central value of the LHCb measurement in eq.~(\ref{eq:RKLHCb}) with muon-specific NP contributions would require~\cite{Alonso:2014csa,Hiller:2014yaa}  
\begin{align}
\delta C_9^{(\prime)\mu}\simeq-1,~~~~~~~~~~~~~\text{or}~~~~~~~~~~~~~\delta C_{10}^{(\prime)\mu}=+1,  \label{eq:RKWC1}
\end{align}
or a suitable combination of the two such as in the leptonic left-handed combination,
\begin{align}
\delta C_L^\mu=\delta C^\mu_9=-\delta C^\mu_{10}=-0.5, \label{eq:RKWC2}
\end{align}
(or its chirally-flipped counterpart $\delta C_L^{\prime\mu}=\delta C^{\prime\mu}_9=-\delta C^{\prime\mu}_{10}$). Explaining the signal instead with electron-specific contributions would require the replacements $\delta C^e\simeq -\delta C^\mu$ in the scenarios above.

The dependence of the $\bar B\to \bar K^*\ell\ell$ rate on the Wilson coefficients is more involved due to the interplay between different helicity amplitudes in the rate. For instance one can express it as
\begin{align}
\frac{d\Gamma_{\bar K^*}}{dq^2}=\frac{d\Gamma_\perp}{dq^2}+\frac{d\Gamma_0}{dq^2},
\end{align}
where $\Gamma_0$ ($\Gamma_\perp$) corresponds to the decay rate into longitudinally or transversally polarized $\bar K^*$, and we define $F_L=\Gamma_0/\Gamma_{\bar K^*}$ as the longitudinal polarization fraction in the decay. 
Expanding the $\bar B\to \bar K^*\ell\ell$ rates around the massless limit of the lepton, one obtains
\begin{align}
\frac{d\Gamma_0}{dq^2}&=\mathcal N_{K^*0}|\vec k|^3V_0(q^2)^2\left(\left|C_{10}^\ell-C_{10}^{\prime\ell}\right|^2+\left|C_9^\ell-C_9^{\prime\ell}+\frac{2m_b}{m_B}C_7\frac{T_0(q^2)}{V_0(q^2)}-8\pi^2 h_{K^*0}\right|^2\right)+\mathcal O\left(\frac{m_\ell^2}{q^2}\right),\label{eq:dG0}\\
\frac{d\Gamma_\perp}{dq^2}&=\mathcal N_{K^*\perp}|\vec k|q^2 V_-(q^2)^2\left(\left|C_{10}^\ell\right|^2+\left|C_9^{\prime\ell}\right|^2+\left|C_{10}^{\prime\ell}\right|^2+\left|C_9^\ell+\frac{2m_bm_B}{q^2}C_7\frac{T_-(q^2)}{V_-(q^2)}-8\pi^2 h_{K^*\perp}\right|^2\right)\nonumber\\
&+\mathcal O\left(\frac{m_\ell^2}{q^2}\right)+\mathcal O\left(\frac{\Lambda}{m_b}\right).\label{eq:dG0perp}
\end{align}
In this formula $\mathcal N_{K^*0,\perp}$ are dimensionful constants, $V_{0,-}(q^2)$ and $T_{0,-}(q^2)$ are form factors in the helicity basis~\cite{Jager:2012uw} and $h_{K^*0,\perp}$ describe the contributions from the four-quark and chromomagnetic operators much like $h_K$ above. Furthermore, we have neglected the hadronic matrix elements giving the leading contributions in the SM to decays into positively polarized $K^*$ (e.g. the form factors $V_+(q^2)$ and $T_+(q^2)$) because, in the large-recoil region (low $q^2$), they are suppressed by  $\mathcal O(\Lambda/m_b)$~\cite{Jager:2012uw}. In the SM these corrections, as well as, in general, the hadronic uncertainties, largely cancel in the $R_{K^*}$ ratios, formally appearing as  $\mathcal O(m_\mu^2/q^2\times\Lambda/m_b)$ terms that will be systematically included in our numerical analysis.

\begin{figure}[h]
\begin{tabular}{cc}
  \includegraphics[width=90mm]{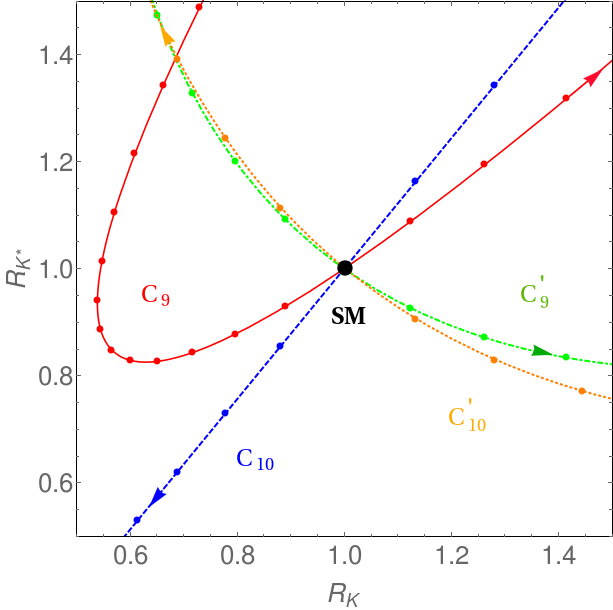}
\end{tabular}
\caption{$R_K$ and $R_{K^*}$ (in the $[1.1,~6]$ GeV$^2$ bin) parametric dependence on one Wilson coefficient at a time, for NP affecting only the muonic coefficients, where the nodes indicate steps of $\Delta(\delta  C^\mu)=+0.5$ from the SM point and in the direction of the arrows. The red solid line shows the dependence on $\delta C_9^{\mu}$, dashed blue line on $\delta C_{10}^{\mu}$, green dot-dashed on $\delta C_9^{\prime\mu}$ and orange dotted on $\delta C_{10}^{\prime \mu}$. 
\label{fig:parametric}}
\end{figure}

The longitudinal contribution to the rate, eq.~(\ref{eq:dG0}), is similar to the $B\to K\ell\ell$ one except that the chirally flipped operators interfere with the SM with a relative minus sign due to the different transformations under parity of the $B\to K$ and $B\to K^*$ hadronic matrix elements. In the transversal polarization, the interference of the chirally flipped operators with the SM is suppressed by the neglected $\Lambda/m_b$ terms in eq.~(\ref{eq:dG0perp}), so that their contributions will always increase $\Gamma_\perp$. Any scenario explaining the deficit in $R_K$ via a destructive interference with the SM in eq.~(\ref{eq:BtoKllrate}) with (small) negative values of $C^\prime_{9,10}$, will necessarily produce a \textit{surplus} in $R_{K^*}$.

Another interesting feature of the transversal rate is the destructive interference that occurs in the SM between $\mathcal O_9$ and $\mathcal O_7$ ($C_7(m_b)=-0.333$) whose contribution is the dominant one at low $q^2$ because of the pole of the virtual photon exchanged with the lepton pair. A negative contribution to $C_9^\mu$ would then \textit{increase} $\Gamma_\perp$,  partially compensating for the deficit that the very same contribution would produce in $\Gamma_0$. Thus, a $C_9$ scenario is not as efficient at reducing $R_{K*}$ at low $q^2$ as it is in $R_K$, or as compared to a contribution from $\mathcal O_{10}$, which would coherently reduce the three decay rates $\Gamma_K$, $\Gamma_0$, and $\Gamma_\perp$.~\footnote{In this paper we will not consider new physics in $C_7^{(\prime)}$ that could contribute through these interference terms. These Wilson coefficients are very tightly constrained by the decay $B\to X_s\gamma$, the time dependent $CP$-asymmetry of $B\to K^*\gamma$ and the angular analysis of $B\to K^*ee$ a very low $q^2$~\cite{Jager:2014rwa,Aaij:2015dea,Paul:2016urs}. } 

In Fig.~\ref{fig:parametric} we show the parametric dependence of $R_K$ and $R_{K^*}$ (in the $[1.1,~6]$ GeV$^2$ bin) on one Wilson coefficient. The nodes indicate steps of $\Delta C^\mu=+0.5$ from the SM point and in the direction of the arrows, illustrating the patterns discussed in this section.

\section{Predictions in the Standard Model and in selected new-physics scenarios}

\begin{figure}[h]
\begin{tabular}{cc}
  \includegraphics[width=83mm]{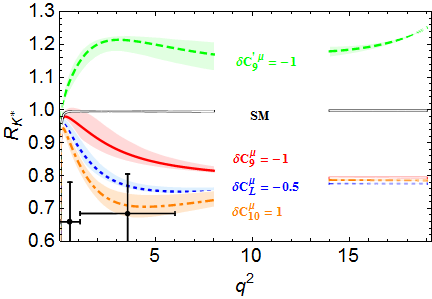}\hspace{0.2cm} &  \includegraphics[width=85mm]{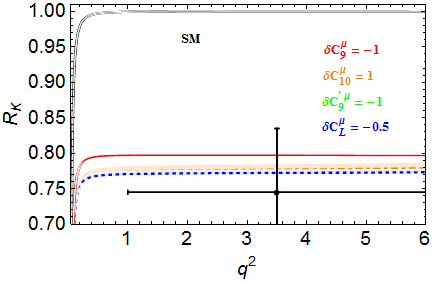} 
\end{tabular}
\caption{Results for $R_{K^*}$ and $R_K$ in the SM and various NP scenarios as a function of the invariant mass of the lepton pair, $q^2$. Solid gray line corresponds to the SM, solid red line to the scenario in which $\delta C_9^{\mu}=-1$, dot-dashed orange line to $\delta C_{10}^{\mu}=1$, dashed green line to $\delta C_9^{\prime\mu}=-1$ and blue dotted line to $\delta C_L^\mu=-0.5$. The shadings around each curve indicates our estimate of the hadronic uncertainties (see main text). We overlay the experimental LHCb results shown in black as points with error bars.
\label{fig:diffdecays}}
\end{figure}

\begin{table}[h]
\centering
\caption{Results for binned observables in the SM and different NP scenarios using the approach described in ref.~\cite{Jager:2014rwa}. The errors are obtained with the gaussian method described in ref.~\cite{Jager:2014rwa} for the distributions of the hadronic parameters. The experimental results have been rounded conservatively by taking the larger side of the statistical error and adding the systematic in quadrature, and neglecting correlations. We show also predictions for the high $q^2$ bin.\label{tab:preds}}
\begin{tabular}{|c|c|c|c|c|c|c|}
\hline
Obs.&Expt.&SM&$\delta C_L^{\mu}=-0.5$& $\delta C_9^\mu=-1$&$\delta C_{10}^\mu=1$&$\delta C_9^{\prime\mu}=-1$\\
\hline
$R_K~[1,~6]$ GeV$^2$&$0.745\pm0.090$&$1.0004^{+0.0008}_{-0.0007}$&$0.773^{+0.003}_{-0.003}$&$0.797^{+0.002}_{-0.002}$&$0.778^{+0.007}_{-0.007}$&$0.796^{+0.002}_{-0.002}$\\
$R_{K^*}~[0.045,~1.1]$ GeV$^2$&$0.66\pm0.12$&$0.920^{+0.007}_{-0.006}$&$0.88^{+0.01}_{-0.02}$&$0.91^{+0.01}_{-0.02}$&$0.862^{+0.016}_{-0.011}$&$0.98^{+0.03}_{-0.03}$\\
$R_{K^*}~[1.1,~6]$ GeV$^2$&$0.685\pm0.120$&$0.996^{+0.002}_{-0.002}$&$0.78^{+0.02}_{-0.01}$&$0.87^{+0.04}_{-0.03}$&$0.73^{+0.03}_{-0.04}$&$1.20^{+0.02}_{-0.03}$\\
$R_{K^*}~[15,~19]$ GeV$^2$&$-$&$0.998^{+0.001}_{-0.001}$&$0.776^{+0.002}_{-0.002}$&$0.793^{+0.001}_{-0.001}$&$0.787^{+0.004}_{-0.004}$&$1.204^{+0.007}_{-0.008}$\\
\hline
\end{tabular}
\end{table}

In Fig.~\ref{fig:diffdecays} and Tab.~\ref{tab:preds} we present predictions for the differential and binned lepton-universality ratios in the SM and in various benchmark NP scenarios of interest, compared to the experimental measurements by the LHCb. The NP scenarios coupling selectively to the electrons produce very similar results to those of the muons replacing $\delta C_i^e\simeq-\delta C_i^\mu$. The calculations are performed following ref.~\cite{Jager:2014rwa} and where the errors are obtained with the gaussian method described there for the distributions of the hadronic parameters. The central values and uncertainties for the inputs are listed in Table II of ref.~\cite{Jager:2014rwa} except for the parametrizations describing the corrections to the heavy-quark limit, whose functional form and numerical values for the error intervals of the parameters are described in Sec. II-A of this reference.   For $R_K$ we extend to $B\to K\ell\ell$ decays the parametrization of power corrections put forward for $B\to K^*\ell\ell$ in ref.~\cite{Jager:2012uw} and use input from light-cone sum rules~\cite{Khodjamirian:2012rm} to fix the only form factor appearing in the heavy-quark limit~\cite{Beneke:2000wa} and to estimate the effects of the soft-gluon exchange in the charm-contribution to the decay (see~\cite{Jager:2012uw,Jager:2014rwa} for details). Uncertainties for the high $q^2$ region account only for the errors of the lattice calculation of the form factors~\cite{Horgan:2013pva} and we do not attempt to quantify the uncertainties of duality violations to the operator-product expansion treatment of the charm~\cite{Grinstein:2004vb} implemented in our analysis. Let us stress again that in the SM these uncertainties almost cancel in the lepton universality ratios, as shown in our predictions in Tab.~\ref{tab:preds}, and are negligible as advertised. This is not longer true in presence of lepton-specific NP contributions, especially for the case of the $R_{K^*}$.~\footnote{It is interesting to note, though, that effects in $R_{K^{(*)}}$ of lepton-specific contributions to $\delta C_9^\ell$ should be enhanced around the mass of resonances (charmoniums at high-$q^2$ or the $\phi$ at low-$q^2$) because of the interference effects.} It is important to stress that radiative electromagnetic corrections, which break lepton universality and can reach the $\gtrsim2\%$ level at the level of the measured ratios~\cite{Bordone:2016gaq}, have not been included in our predictions.

The dependence on the Wilson coefficients of interest of $R_K$ and $R_{K^*}$ in the bins $[1,~6]$ GeV$^2$ and $[1.1,~6]$ GeV$^2$, respectively, as functions of the Wilson coefficients of the four semileptonic operators can be expressed as 
\begin{align}
\frac{\Gamma_K}{\Gamma_K^{\rm SM}}=&\left(2.9438~\left(|C_9+C_9^\prime|^2+|C_{10}+C_{10}^\prime|^2\right)-2{\rm Re}\left[(C_9+C_9^\prime)(0.8152+i~0.0892)\right]+0.2298\right)~10^{-2},\nonumber\\
\frac{\Gamma_{K^*}}{\Gamma_{K^*}^{\rm SM}}=&\left(2.420\left(|C_9-C_9^\prime|^2+|C_{10}-C_{10}^\prime|^2\right)-2{\rm Re}\left[(C_9-C_9^\prime)(2.021+i~0.188)\right]+1.710\right.\nonumber\\
&\left.+1.166\left(|C_9|^2+|C_{10}|^2+|C_9^\prime|^2+|C_{10}^\prime|^2\right)-2{\rm Re}\left[C_9(5.255+i~0.239)\right]+29.948\right)~10^{-2},\label{eq:numfoms}
\end{align}  
where the numerical coefficients have been obtained using ref.~\cite{Jager:2014rwa} at a scale $\mu=4.65$ GeV. These formulas give results on the ratios which are accurate at the $\sim1\%$ level (which is better than the hadronic uncertainties)
when compared to the exact numerical calculation and can be directly used to obtain predictions for NP defects in muons or in electrons. 

In light of the discussion above and given the \textit{deficits} observed in both $R_K$ and $R_{K^*}$ we conclude that the primed operators $\mathcal{O}_{9,10}^\prime$ are disfavored by the data. Henceforth we will focus only on the analysis of the operators  $\mathcal{O}_{9}$ and  $\mathcal{O}_{10}$.

\section{Fits to $R_{K^{(*)}}$ and other $b\to s\ell\ell$ data}

In this section we assess the compatibility with the SM of
lepton-universality-violating and -conserving rare $B$-decay measurements,
and perform frequentist fits to (NP values of) Wilson coefficients
in the weak Hamiltonian. In all cases, theoretical uncertainties
are modeled by adding a ``theory term'' to the $\chi^2$,
\begin{align}
\tilde \chi^2(\vec C,~\vec y)=\chi^2_{\rm exp}(\vec C,~\vec y)+\chi^2_{\rm th}(\vec y).
\end{align}        
Here $\chi^2_{\rm exp}(\vec C,~\vec y)$ is constructed
in the usual way out of the experimental measurements considered,
and the theoretical expressions for them in terms of a vector of
Wilson coefficients $\vec C$ included in a given fit, and
the 27-component vector $\vec y$  of hadronic parameters
that enter in the model-independent description
of the $B\to K^*\ell\ell$~\cite{Jager:2012uw,Jager:2014rwa}. The theory term is taken in gaussian form
\begin{align}
\chi^2_{\rm th}(\vec y)=\sum_i\left(\frac{y_i-\bar y_i}{\delta y_i}\right)^2,\label{eq:chi2th}
\end{align}
where $\bar y_i$ are the central values of the theory parameters
and $\delta y_i$ their estimated uncertainties, which we assume to be fully uncorrelated~\cite{Jager:2014rwa}.~\footnote{Theoretical uncertainties for $R_K$ are always very small 
compared to the experimental errors in all scenarios considered (see Fig.~\ref{fig:diffdecays} and Tab.~\ref{tab:preds}) and are neglected in our analysis.}

Once it is in this form, we treat the $y_i$ as nuisance parameters
and construct a profile $\chi^2$ that depends on the Wilson coefficients
only,
\begin{align}
\chi^2(\vec C)=\underset{\vec{y}}{\rm min}\,\tilde\chi^2(\vec{C},\vec{y}),
\end{align}
from which we determine frequentist confidence level intervals and
contours for the Wilson coefficients. The construction here is
nothing but the Rfit method \cite{Hocker:2001xe} with gaussian theory errors
instead of parameter ranges.

\subsection{Fits only to $R_K $ and $R_{K^*}$}
\label{sec:fitRKRKs}

We begin by considering $R_K $ and $R_{K^*}$ alone.
Setting the Wilson coefficients to their SM values, we obtain
$\chi^2_{\rm min,SM} = 19.51$ 3 degrees of freedom (d.o.f.), corresponding to a $p$-value
of $2.1 \times 10^{-4}$ or a $3.7 \sigma$  deviation. 
It must be stressed that the  significance of this
evidence for new physics in lepton-universality ratios alone
is fully dominated by experimental statistical errors,
with hadronic uncertainties (the $y_i$-parameters) playing almost no role,
due to the exceptional cleanness of the observables.
Nonetheless, in our fit we include the theoretical errors
of $R_{K^*}$ at low $q^2$ and investigate below
the robustness of our fits against variations of the theoretical error ranges.

\begin{figure}[h]
\begin{tabular}{ccc}
  &  \includegraphics[width=70mm]{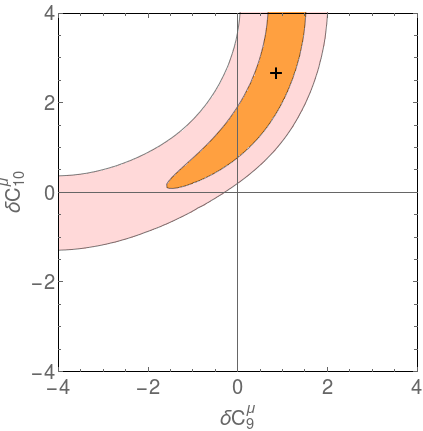} &\\
  \includegraphics[width=50mm]{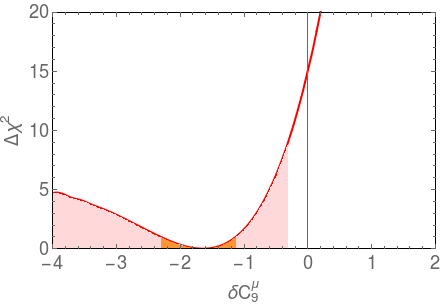} &  \includegraphics[width=50mm]{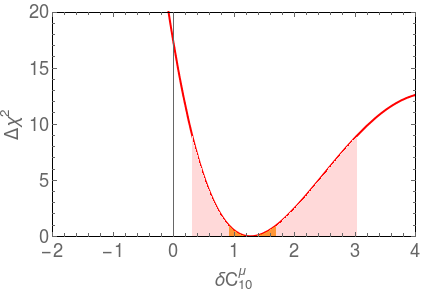} &    \includegraphics[width=50mm]{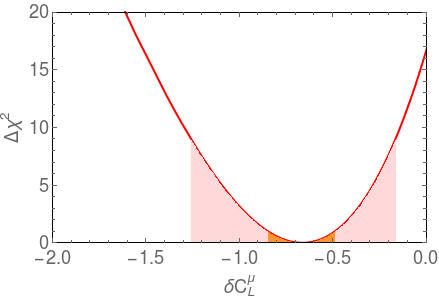}
\end{tabular}
\caption{The top panel shows the result of a best fit to a model that includes muon-specific NP Wilson coefficients $\delta C_9^\mu$ and $\delta C_{10}^\mu$. The cross indicates the position of the minimum. The first two graphs in the bottom row give the $\chi^2$ distribution projected onto each Wilson coefficient, while the third one is projected onto the difference $\delta C_L^\mu=\delta C_9^\mu-\delta C_{10}^\mu$.
Ranges in orange and light red correspond to 1σ and 3σ intervals of Wilson coefficients, respectively. ($\Delta\chi^2=1 (9)$ for $1\sigma(3 \sigma)$ in the 1-parameter cases,
$\Delta\chi^2=2.3 (11.83)$ for $1\sigma(3 \sigma)$
  in the 2-parameter fit.) 
\label{fig:fitRKRKs}}
\end{figure}

We next show that a good fit is obtained by allowing for
lepton-specific  contributions $\delta C_9^\ell$ and $\delta C_{10}^\ell$.
In Table \ref{tab:fitRKRKs}
and Figure \ref{fig:fitRKRKs} we display the results of one- and two-dimensional
fits of muon-specific NP Wilson coefficient values $\delta C_9^\mu$ and
$\delta C_{10}^\mu$.

\begin{table}[h]
\centering
\caption{ 
    Best fit values, goodness of fit, $p$-value, SM exclusion level (pull), and
  confidence intervals for fits of single or pairs of
  Wilson coefficients, to $R_{K}$ and $R_{K^*}$ data.
For the one-dimensional case, we show that the 1$\sigma$ and 3$\sigma$ confidence intervals, while for the two-dimensional case we show the
  one-sigma intervals for the two parameters instead.
   \label{tab:fitRKRKs} }
{\normalsize
  \begin{tabular}{|c|c|c|c|c|c|c|}
\hline
Coeff. & best fit & $\chi^2_{\rm min}$ & $p$-value & SM exclusion [$\sigma$] & 1$\sigma$ range & 3$\sigma$ range\\\cline{1-5}
\hline
$\delta C_9^{\mu}$ & -1.64  & 4.52 & 0.104 & 3.87 &[-2.31,-1.13]  & [$<$-4,~-0.31] \\

$\delta C_{10}^{\mu}$ & 1.27  &  2.24 & 0.326 &4.15 & [0.91,1.70] & [0.31,3.04] \\

$\delta C_L^{\mu}$ & -0.66 & 2.93 & 0.231 & 4.07& [-0.85,-0.49]  & [-1.26,-0.16] \\
\hline
Coeff. & best fit & $\chi^2_{\rm min}$ & $p$-value & SM exclusion [$\sigma$] & \multicolumn{2}{c|}{parameter ranges}\\\cline{1-5}
\hline
$(\delta C_9^{\mu},\delta C_{10}^{\rm \mu})$ & (0.85, 2.69) & 1.99 & 0.158 & 3.78
   & $C_9^\mu \in$ [-0.71,~1.38] & $C_{10}^\mu \in$ [0.61,~$>$4] \\\cline{1-5}
\hline
\end{tabular}
}
\end{table}

Good to excellent fits are
obtained in one-parameter scenarios where only  $\delta C_{10}^\mu$,
$\delta C_L^\mu = \delta C_9^\mu = - \delta C_{10}^\mu$, or
$\delta C_9^\mu$ are nonzero, as well as in the two-parameter scenario.
The largest $p$-value (best fit) is obtained in the pure $C_{10}^\mu$-case,
though the differences are not large.
We next consider $\Delta \chi^2 = \chi^2 - \chi^2_{\rm min}$ and use it
to compute the significance at which the SM point (origin)
is excluded (often called ``pull''), and
to construct frequentist confidence regions.
In each of the four scenarios, the SM point (origin) is excluded at $4\sigma$
confidence, or close to it (see Table~\ref{tab:fitRKRKs}, where we also
display best-fit values and confidence intervals for the parameters
(in the two-dimensional case, these are obtained by minimising $\Delta \chi^2$
over the other parameter).

Instead of considering muon-specific effects, we could have assumed
an electron-specific effect, or a combination. In the former case,
essentially the signs of the fitted Wilson coefficients are reversed,
if only the LUV observables are considered.

\subsection{Fits to $R_K$, $R_{K^*}$ and $B_s\to \mu\mu$}

\label{sec:fitRKRKsBsmumu}

We now add $\overline{\rm BR}(B_s\to\mu\mu)$ to the data set.\footnote{The overline refers to the fact that 
the experiments access the time-integrated branching ratio,
which depends on the details of $B_s \bar B_s$ mixing~\cite{DeBruyn:2012wk}.}
It is theoretically very clean
with  NNLO QCD and NLO electroweak corrections
known~\cite{Bobeth:2013uxa},
and the sole
hadronic parameter, the decay constant $f_{B_s}$, having been precisely
computed by different lattice QCD collaborations~\cite{Aoki:2016frl}.
To simplify the fit, we consider the ratio
\begin{align}
 R=\frac{\overline{\rm BR}(B_s\to\mu\mu)}{\overline{\rm BR}(B_s\to\mu\mu)^{\rm SM}}=\left|\frac{C_{10}^\mu 
 }{C_{10}^{\rm SM}}\right|^2 ,
\end{align}
where we have neglected the contributions of scalar operators. Among
the set $(C_9^{\ell}, C_{10}^\ell)$, this branching fraction only depends
on the coefficient $C_{10}^\mu$, such that it is natural to add it to
the fit of muon-specific Wilson coefficients.
As experimental value for $R$ we employ the combination obtained in ref.~\cite{Altmannshofer:2017wqy}, $R^{\rm exp}=0.83(16)$, 
where the results from
CMS and LHCb including run I and run II data
have been averaged. The error includes, in
quadrature, the theory uncertainty on the SM rate,
which is small compared to the experimental ones.

\begin{figure}[h]
\begin{tabular}{cc}
  \includegraphics[width=80mm]{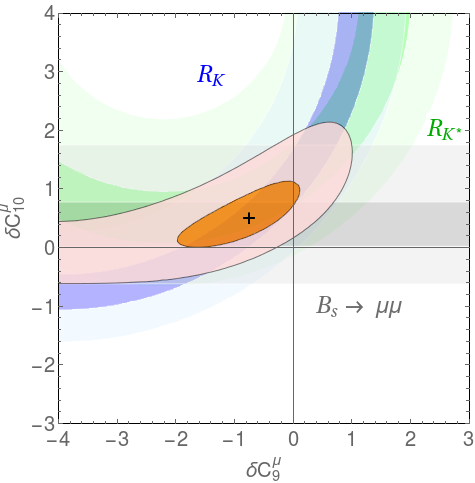}
\end{tabular}
\caption{Contours at $1\sigma$ and $3\sigma$ level in the $(\delta C_9^\mu,~\delta C_{10}^\mu)$ plane, in solid lines and orange and light-red colors, for the fit to $R_K$, $R_{K^*}$ 
and $\overline{\rm BR}(B_s\to \mu\mu)$. We also show the $1\sigma$ and $3\sigma$ constraints  given individually by $R_K$, $R_{K^*}$ in the [1.1,6] GeV$^2$ bin and $\overline{\rm BR}(B_s\to \mu\mu)$ using
blue, green and gray contours, respectively. The cross indicates the position of the minimum.  
\label{fig:RKRKsBsmumu2D}}
\end{figure}

Including $R$ increases the SM $p$-value marginally
to $3.7~10^{-4}$ (3.56$\sigma$).
We next
perform the same fits as in the previous subsection, but to the extended
data set. The results are shown in Tab.~\ref{tab:fitRKRKsBsmumu} and, for the fit of $(\delta C_9^{\mu},\delta C_{10}^{\rm \mu})$ fit, in Fig.~\ref{fig:RKRKsBsmumu2D}.

\begin{table}[h]
\centering
\caption{
  As in Table~\ref{tab:fitRKRKs} but in fits to $R_K$, $R_{K^*}$ and $B_s\to \mu\mu$ data.
 \label{tab:fitRKRKsBsmumu} }
{\normalsize
  \begin{tabular}{|c|c|c|c|c|c|c|}
\hline
Coeff. & best fit & $\chi^2_{\rm min}$ & $p$-value & SM exclusion [$\sigma$] & 1$\sigma$ range & 3$\sigma$ range\\\cline{1-5}
\hline
$\delta C_9^{\mu}$ & -1.64  & 5.65 & 0.130 &3.87 &[-2.31,~-1.12]  & [$<$-4,~-0.31] \\

$\delta C_{10}^{\mu}$ & 0.91  &  4.98 & 0.173& 3.96& [0.66,~1.18] & [0.20,~1.85] \\

$\delta C_L^{\mu}$ & -0.61 & 3.36 & 0.339&4.16 & [-0.78,~-0.46]  & [-1.14,~-0.16] \\
\hline
Coeff. & best fit & $\chi^2_{\rm min}$ & $p$-value & SM exclusion [$\sigma$] & \multicolumn{2}{c|}{parameter ranges}\\\cline{1-5}
\hline
$(\delta C_9^{\mu},\delta C_{10}^{\rm \mu})$ & (-0.76, 0.54) & 3.31 & 0.191& 3.76
   & $C_9^\mu \in$ [-1.50,~-0.16] & $C_{10}^\mu \in$ [0.18,~0.92] \\\cline{1-5}
\hline
\end{tabular}
}
\end{table}
Again, all four scenarios considered provide good fits. The main impact on the two-parameter fit is that the allowed
region is narrowed down considerably, with large positive correlated values
of $\delta C_9^\mu$ and $\delta C_{10}^\mu$ no longer allowed.
We note, in particular, that the combination of
$C_L^\mu = \frac{1}{2} (C_9^\mu - C_{10}^\mu)$ is nonzero at more than
$4\sigma$ significance, and is relatively well determined by the LUV data set
plus $B_s \to \mu \mu$ alone, irrespective of and in contradistinction from
the value of $C_R^\mu = \frac{1}{2} (C_9^\mu + C_{10}^\mu)$, which is consistent
with zero and poorly constrained.

\subsection{Fits to $R_K$, $R_{K^*}$, $B_s\to \mu\mu$ and $B\to K^*\ell\ell$ data}
\label{sec:fitGlobal}

We now  include in the fits all measurements of the angular distribution in
$B\to K^*\mu\mu$ by LHCb, ATLAS, CMS, and Belle in the
low-$q^2$ region $q^2\lesssim6$ GeV$^2$ (except for lepton-universality differences measured by Belle, for which we do include the [4, 8] GeV${}^2$ bin).
The reason for this restriction is that we can then reasonably estimate the size of the hadronic uncertainties. As advertised above, we
will later quantify the robustness of our conclusions with respect
to the size of the theoretical errors. The precise dataset comprises
\begin{itemize}
 \item {\bf LHCb:} The 32 measurements of the $CP$-averaged angular observables $F_L$, $P_1$, $P_2$, $P_3$, $P_4^\prime$, $P_5^\prime$, $P_6^\prime$, $P_8^\prime$ in the bins $[0.1,~0.98]$ GeV$^2$,  $[1.1,~2.5]$ GeV$^2$,  $[2.5,~4]$ GeV$^2$ and $[4,~6]$ GeV$^2$~\cite{Aaij:2015oid} including, inside of each bin, the published correlations among the observables. 
\item {\bf ATLAS:} The 18 measurements of the $CP$-averaged angular observables $F_L$, $P_1$, $P_4^\prime$, $P_5^\prime$, $P_6^\prime$, $P_8^\prime$ in the bins $[0.04,~2]$ GeV$^2$, $[2,~4]$ GeV$^2$, $[4,~6]$ GeV$^2$ reported in Moriond EW 2017~\cite{ATLAS:2017dlm}. Correlations of this data are not public and are neglected.
\item {\bf CMS:} The 6 measurements of the $CP$-averaged angular observables $P_1$ and $P_5^\prime$  in the bins $[1,~2]$ GeV$^2$, $[2,~4.3]$ GeV$^2$, $[4.3,~6]$ GeV$^2$ reported in Moriond EW 2017~\cite{CMSDinardo}.  Correlations of this data are not public and are neglected.
\item {\bf Belle:} The 4 measurements of the $CP$-averaged lepton universality differences of angular observables $Q_4$ and $Q_5$ in the bins $[1,~4]$ GeV$^2$ and $[4,~8]$ GeV$^2$~\cite{Wehle:2016yoi}.  Correlations of this data are not public and are neglected.
\end{itemize}
In addition, we include in the fit the ${\rm BR}(B\to K^*\gamma)$~\cite{Coan:1999kh,Nakao:2004th,Aubert:2009ak,Patrignani:2016xqp,Horiguchi:2017ntw} which provides an important constraint to a hadronic form factor in the fit~\cite{Jager:2014rwa}. Including 
the measurements of $R_K$, $R_{K^*}$ and $\overline{\rm BR}(B_s\to\mu\mu)$, the total number of measurements in the fit
is 65.

The resulting $\chi^2_{\rm min, SM}$-value is 81.1 [65 d.o.f.], corresponding
to a $p$-value of $0.086$. Note that this is considerably larger than
before---adding the many angular observables, the significance of the anomalies
has decreased. Each of the four models we consider provides an excellent fit,
once the full data set is considered.
  At the same time, the significance of the SM exclusion in three
  of the four fits is above $4\sigma$; we
   show fit results in Table \ref{tab:fitallmuonic}.
  \begin{table}[h]
    \centering
\caption{
      Same as Table \ref{tab:fitRKRKsBsmumu}, but with  ${\rm BR}(B\to K^*\gamma)$, the
    $B \to K^* \mu \mu$ angular distribution and $Q_i$ observables
  added to the data set.
\label{tab:fitallmuonic} }
{\normalsize
  \begin{tabular}{|c|c|c|c|c|c|c|}
\hline
Coeff. & best fit & $\chi^2_{\rm min}$ & $p$-value & SM exclusion [$\sigma$] & 1$\sigma$ range & 3$\sigma$ range\\\cline{1-5}
\hline
$\delta C_9^{\mu}$ & -1.37  & 61.98 [64 dof]& 0.548& 4.37 &[-1.70,~-1.03]  & [-2.41,~-0.41] \\

$\delta C_{10}^{\mu}$ & 0.60  &  71.72 [64 dof]& 0.237 & 3.06 & [0.40,~0.82] & [-0.01,~1.28] \\

$\delta C_L^{\mu}$ & -0.59 & 63.62 [64 dof]& 0.490 & 4.18 & [-0.74,~-0.44]  & [-1.05,~-0.16] \\
\hline
Coeff. & best fit & $\chi^2_{\rm min}$ & $p$-value & SM exclusion [$\sigma$] & \multicolumn{2}{c|}{parameter ranges}\\\cline{1-5}
\hline
$(\delta C_9^{\mu},\delta C_{10}^{\rm \mu})$ & (-1.15,~0.28) & 60.33 [63 dof] & 0.572& 4.17
   & $C_9^\mu \in$ [-1.54,~-0.81] & $C_{10}^\mu \in$ [0.06,~0.50] \\\cline{1-5}
\hline
\end{tabular}
}
\end{table}
\begin{figure}[h]
\begin{tabular}{ccc}
  \includegraphics[width=80mm]{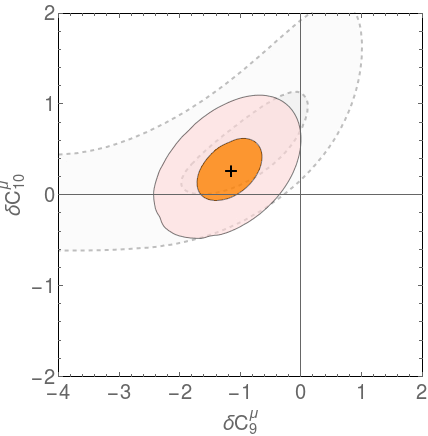}
\end{tabular}
\caption{Contours at $1\sigma$ and $3\sigma$ level in the $(\delta C_9^\mu,~\delta C_{10}^\mu)$ plane, in solid lines and orange and light-red colors, for the fit to LUV data,   
$\overline{\rm BR}(B_s \to \ell \ell)$, ${\rm BR}(B\to K^*\gamma)$, the $B \to K^* \mu \mu$ angular distribution, the $Q_i$ observables, and 
the $B \to K^* e^+ e^-$ angular observables in the ultralow bin. The cross indicates the position of the minimum. Excluding the latter electronic data produces a very similar fit. 
Underneath, and for comparison purposes, we show in gray colors and dashed lines the 1$\sigma$ and  3$\sigma$ regions for the fit to only LUV data and $\overline{\rm BR}(B_s \to \ell \ell)$ 
described in the Sec.~\ref{sec:fitRKRKsBsmumu} and plotted in Fig.~\ref{fig:RKRKsBsmumu2D}. 
  \label{fig:fitallmuonic}}
\end{figure}
The main impact on the fits is to narrow down the $C_9^\mu$ range
further, excluding positive values at high confidence.
Note that the axes of the, approximately elliptic, allowed region in
the two-parameter fit are
nearly aligned with the $C_L^\mu$ and $C_R^\mu$ directions, in such a fashion
that $C_L^\mu$ is again nonzero at more than $4\sigma$ confidence,
while $C_R^\mu$ is more or less consistent with zero,
and much less well determined.

Finally, we consider the measurements of $B \to K^* e^+ e^-$ performed by
LHCb, namely of the branching fraction \cite{Aaij:2013hha}, as well as of four angular
observables \cite{Aaij:2015dea} with a reduced form-factor depencence
similarly to the muonic case. A fit including the resulting 70 measurements is shown in Figure~\ref{fig:fitallmuonic}.
Including or omitting the electron data has only a minimal impact on the fit and statistical tests.

\subsection{Robustness of fit with respect to hadronic uncertainties}

\begin{figure}[h]
\begin{tabular}{cc}
 \includegraphics[width=90mm]{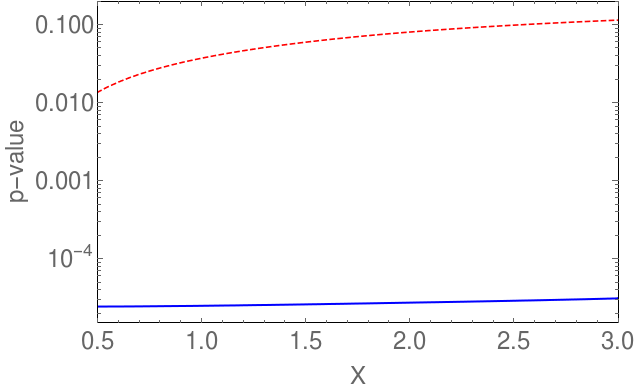} 
\end{tabular}
\caption{\textit{Robustness of fit:} Plot of the $p$-values of the SM as a function of $x$ and in two different fits: Solid (Blue) line represents the fit to the $R_K$ and $R_{K^*}$ ratios and $B_s\to \mu\mu$  
and the dashed (red) line represents the global fit including the $B\to K^*\mu\mu$ angular observables. The variable $x$ is a factor by which we multiply all the uncertainty ranges of the theoretical parameters in eq.~(\ref{eq:chi2th}). 
\label{fig:Robustness}}
\end{figure}

We investigate the robustness of our fits to the theoretical uncertainties of the $B\to K^*\mu\mu$ amplitude, which can affect considerably the predictions of the angular analysis and, therefore, the tensions of the SM with the data. In order to do so, we perform a scan of the variable $x$, that is a factor by which we multiply all the uncertainty ranges of the theoretical parameters in eq.~(\ref{eq:chi2th}), in the range $x\in[0.5,~3]$. At each $x$, we calculate the variation of the $\chi^2_{\rm d.o.f.}$ with respect to $x$ in the SM in the fit to only $R_K$, $R_{K^*}$ and $\overline{\rm BR}(B_s\to\mu\mu)$.~\footnote{We do not modify the value of $f_{B_s}$ which is obtained from the FLAG average of lattice QCD calculations~\cite{Aoki:2016frl}.} The results are shown in Fig.~\ref{fig:Robustness} by the blue solid curve, which demonstrates the stability, with respect to the hadronic uncertainties in the semileptonic decays, of the fits to the lepton-universality ratios. This is just a consequence of the cancellations of hadronic uncertainties in the ratios discussed in previous sections. For illustrative purposes, we compare with the fit in which we also include the 61 measurements of observables in the angular distribution of $B\to K^*\mu\mu$ and the radiative $B\to K^*\gamma$ decay, which shows a much stronger sensitivity, as expected. A more careful treatment of hadronic uncertainties and their impact in the interpretation and analysis of the robustness of
the global fits is left for future studies.

\subsection{Beyond muon-specific lepton nonuniversality}
\begin{figure}[h]
\begin{tabular}{ccc}
  \includegraphics[width=70mm]{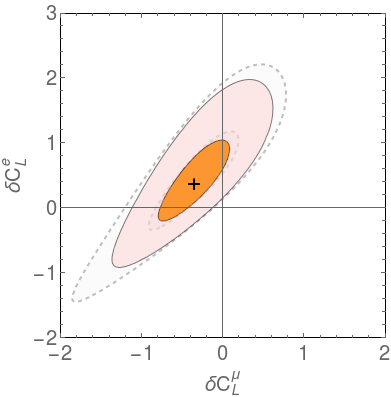}
& \hspace{0.5cm} &
 \includegraphics[width=70mm]{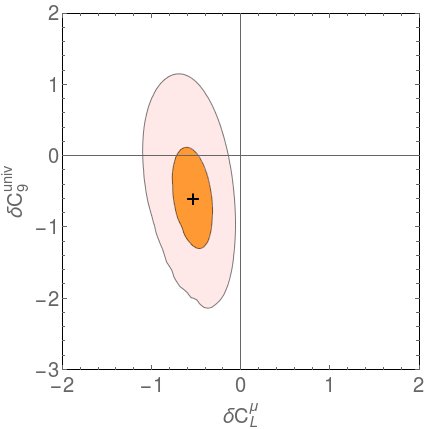} 
\end{tabular}
\caption{\textit{Left panel:} Two-parameter fit to $C_L^\mu$ and $C_L^e$ where the gray regions and dashed contours
include only $R_K$, $R_{K^*}$ and $B_s\to\mu\mu$. The overlaid orange and light-red colored regions, enclosed by solid lines, represent the 1$\sigma$ and 3$\sigma$ bounds 
including the full data set. \textit{Right panel:}
Two-parameter fit to $C_9^e=C_9^\mu=C_9^{\rm univ}$ and $C_L^\mu$ where we only show the 1$\sigma$ and 3$\sigma$ regions for the fit to the full data set. 
The crosses indicate the positions of the minima. 
  \label{fig:beyondmuon}}
\end{figure}

As already noted, our choice to focus on new physics in muonic Wilson
coefficients $C_9^\mu$, $C_{10}^\mu$ was far from mandatory. As an example
of a more general scenario that can accommodate the data, in the left panel of Figure
\ref{fig:beyondmuon}, we show a two-parameter fits of $C_L^\mu$ and $C_L^e$,
to LUV data and $B_s \to \mu^+ \mu^-$
and the full data set. In both cases, a high
degree of degeneracy is seen, and pure $C_L^e$ fits the data nearly as
well as $C_L^\mu$. 

In the right panel of Figure \ref{fig:beyondmuon} we consider instead a two-parameter scenario
with a muon-specific Wilson coefficient $C_L^\mu$ and a lepton-flavour-universal
coefficient $C_9^\mu = C_9^e = C_9$.
Such a universal coefficient  generically expected to be
generated in BSM scenario at the one-loop level through penguin diagrams.
One particular realisation is the ``charming BSM scenario'' of
\cite{Jager:2017gal}. In fact,
this ``semi-universal'' scenario fits the data minimally better
the pure-$C_L^\mu$  ($p = 62 \%$ versus $p = 60 \%$). This preference for
an extra $C_9$ effect can be understood as a
consequence of the $B \to K^* \mu^+ \mu^-$
angular distribution measurements (particularly the $P_5'$ term); because the
$B \to K^* e^+ e^-$ measurements are much less precise and only cover
$q^2 < 1.12$ GeV${}^2$, the data do not require this to be muon-specific.
The preference for an extra, possibly universal, $C_9^{\rm univ}$ effect would increase with
more optimistic error estimates on $B$-decay form factors that are sometimes
made in the literature.

\section{Precision probes of a lepton-nonuniversal $C_{10}$}

As shown above, low values of $R_K^*$ at small dilepton mass,
as suggested by the new
LHCb measurements, require a modification of $C_{10}^\mu$. However, the
value of $C_{10}^\mu$ is poorly determined by the global fit:
while the combination $C_L^\mu = (C_9^\mu - C_{10}^\mu)/2$ is already
well determined and significantly different from zero,
the combination $C_R^\mu = (C_9^\mu + C_{10}^\mu)/2$ is poorly constrained.
Breaking the degeneracy between $C_9^\mu$ and $C_{10}^\mu$
will be a key requirement for identifying the dynamics underlying
the NP signals. Currently, the most precise selective probe of the
coefficient $C_{10}^\mu$ is the  $B_s \to \mu^+ \mu^-$ branching fraction,
but this suffers from low statistics. In this section we
describe observables which are selectively sensitive
to a lepton-nonuniversal $C_{10}$ effect at a (theoretical)
precision better than the percent level. They are sensitive to
$C_{10}^\mu$ as well as $C_{10}^e$.
  
The starting point is the observation that
the angular distribution in $B \to K^* \ell^+ \ell^-$
contains a single term proportional to $C_{10}$, commonly
called $I_6$, due to vector-axial-vector interference
and responsible for the lepton forward-backward asymmetry.
In terms of the helicity amplitudes,
\begin{equation}
  I_6^{(\ell)} = N C_{10}^\ell  q^2 \beta_\ell^2(q^2)|\vec k| \left(
      {\rm Re} [H^{(\ell)}_{V-}(q^2)] V_-(q^2)
       + {\rm Re} [H^{(\ell)}_{V+}(q^2) \frac{H^{(\ell)}_{A+}(q^2)}{C_{10}^\ell} ] 
  \right) ,
\end{equation}
where $N$ is a $q^2$-independent normalisation, and
$\beta_\ell^2 = 1 - 4 m_{\ell}^2/q^2$ is the velocity-squared of the
lepton in the dilepton rest frame (equal to one for electrons,
for practical purposes).
As shown in  \cite{Jager:2012uw,Jager:2014rwa}, the amplitude $H_{V+}$ is
suppressed relative to $H_{V-}$ by $q^2 \Lambda/m_B^3$
at small $q^2$-values. Moreover, if the disfavoured Wilson coefficients
$C_9^{\prime\ell}$ and $C_{10}^{\prime\ell}$ are neglected,
$H_{A+}^{(\ell)} = C_{10}^\ell V_+$ is
itself suppressed by one power of $\Lambda/m_B$ in the heavy-quark limit.
The second term hence gives at most a one-percent contribution below $q^2 = 1$
GeV${}^2$ (but will be kept in our
numerics).

Following \cite{Jager:2014rwa}, we define
\begin{equation}
  \label{eq:R6}
  R_6[a,b] =\frac{\int_a^b \Sigma_6^{\mu} dq^2}{\int_a^b \Sigma_6^{e} dq^2}
  \approx
  \frac{C_{10}^{\mu}}{C_{10}^{e}} \times
  \frac{\int_a^b |\vec k| q^2 \beta_\mu^2
      \, {\rm Re} [H_{V-}^{(\mu)}(q^2)] V_-(q^2)}
  {\int_a^b |\vec k| q^2 \, {\rm Re} [H_{V-}^{(e)}(q^2)] V_-(q^2)} ,
\end{equation}
where $\Sigma_6$ is the CP-average of $I_6$. The
sensitivity of $R_6$ to deviations from $C_{10}^\mu/C_{10}^e$ from one
is evident from the direct proportionality exhibited
in (\ref{eq:R6}).
It was noted in \cite{Jager:2014rwa} but not quantified.
Moreover, the constant of proportionality---the ratio of integrals---
differs from one only through the exactly known velocity-factor $\beta_\mu^2$
and possible lepton-flavour-universality violation in $H_{V-}$. 
The only source of such lepton-flavour-nonuniversality would
be a lepton-nonuniversal $C_9^\ell$.
Due to the dominance of the photon pole at small $q^2$,
for $|C_9^{\rm NP}|\lesssim1$
the resulting modification of $H_{V-}$ is at the
$5\%$ level at $q^2 = 1$ GeV${}^2$ and drops roughly linearly with $q^2$ below.
Since $\Sigma_6$ is rather flat in $q^2$ below 1 GeV${}^2$ (see Figure 6 in
\cite{Jager:2012uw}), we conclude that the
impact of NP effects in $C_9^{\rm NP}$ in $R_6[0.045, 1.1]$ GeV$^2$ is at most
$2-3 \%$. Moreover, this small effect is calculable with ${\cal O}(10-20\%)$
relative accuracy in the heavy-quark expansion.
As a result, if
$\delta C_{10}^e = 0$ then $R_6$ can theoretically determine $\delta C_{10}^\mu$
to an (absolute) accuracy of about $0.13$ irrespective
of the value of $\delta C_9^\ell$. Below that level, $R_6$ still probes
a combination of (mainly) $\delta C_{10}^{\ell}$ and $\delta C_9^{\ell}$,
which may still be useful in disentangling the two
in the context of a global fit.

\begin{figure}[t]
\begin{tabular}{cc}
  \includegraphics[width=83mm]{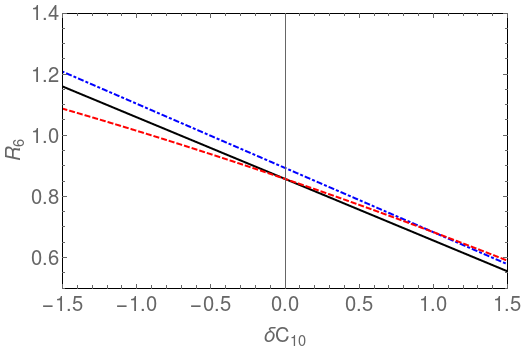}\hspace{0.2cm} &  \includegraphics[width=82mm]{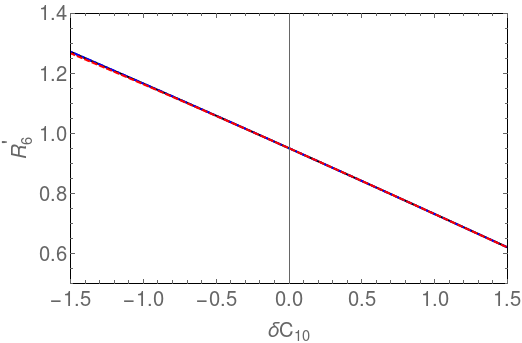} 
\end{tabular}
  \caption{
    \label{fig:R6}
    Studies of sensitivity of the new precision probes $R_6$, Eq.~(\ref{eq:R6}), and $R_6^\prime$, Eq.~(\ref{eq:R6p}), to NP scenarios.
    \textit{Left panel:} $R_6[0.045, 1.1]$ GeV${}^2$ as a function of $\delta C_{10}$ for different values
    of $\delta C_9$: $\delta C_9 = 0$ (black solid),
    $\delta C_9 = - \delta C_{10}$ (red dashed),
    $\delta C_9 = -1$ (blue dot-dashed).
   \textit{Right panel:} Same for $\ R_6^\prime[0.045, 1.1]$ GeV${}^2$.}
\end{figure}

Figure \ref{fig:R6} shows $R_6$
for the bin $[0.045, 1.1]$ GeV${}^2$ as a function of $C_{10}^\mu$ 
for three scenarios differing in $\delta C_9$.
\begin{table}[t]
  \begin{center}
\begin{tabular}{|c|c|c|c|c|c|c|}
\hline
Obs.&SM&$\delta C_L^{\mu}=-0.5$& $\delta C_9^\mu=-1$&$\delta C_{10}^\mu=1$&$\delta C_9^{\prime\mu}=-1$\\
\hline
      $R_6[0.045, 1.1]$ GeV${}^2$ & $0.8571^{+0.0021}_{-0.0012}$&$0.7721^{+0.0006}_{-0.0014}$& $0.8930^{+0.0016}_{-0.0047}$  &$0.6555^{+0.0016}_{-0.0007}$  & $0.8570^{+0.0021}_{-0.0011}$\\
      $\bar R_6[0.045, 1.1]$  GeV${}^2$ & $0.932^{+0.007}_{-0.005}$ & $0.877^{+0.010}_{-0.016}$&$0.985^{+0.008}_{-0.019}$  &$0.761^{+0.010}_{-0.013}$ &$0.877^{+0.030}_{-0.018}$ \\
      $R'_6[0.034, 1.1]$  GeV${}^2$ &$0.9494^{+0.0005}_{-0.0006}$& $0.8403^{+0.0006}_{-0.0014}$  & $0.9494^{+0.0004}_{-0.0014}$  & $0.7300^{+0.0008}_{-0.0013}$& $0.948^{+0.003}_{-0.003}$\\
\hline
      $Q_2[0.045, 1.1]$  GeV${}^2$ &$-0.0081^{+0.0012}_{-0.0005}$ &$-0.026^{+0.004}_{-0.002}$ & $-0.0081^{+0.0012}_{-0.0005}$& $-0.043^{+0.006}_{-0.003}$ &$-0.008^{+0.001}_{-0.001}$ \\
\hline
\end{tabular}
  \end{center}
  \caption{\label{tab:R6} Predictions for the $C_{10}$
    analyzers $R_6$, $\bar R_6$, and $R_6'$ for the four benchmark scenarios
    consistent with the $R_K$ measurement. The observable $Q_2$ is
  also shown, for comparison.}
\end{table}
The sensitivity to $\delta C_{10}^\mu$ is evident from the slope and
the narrowness of the bands in the figure, while the near-degeneracy
of the three bands illustrates the insensitivity to $\delta C_9^\mu$.
Numerical values are given in Table \ref{tab:R6}. They corroborate
the preceding discussion and show that $R_6$ can clearly discriminate
between the two benchmarks that remain viable in the light of
the $R_{K^*}$ measurements.

From an experimental perspective, $R_6$ may benefit from cancellations
of certain systematics in the ratio, and from the fact that both numerator
and denominator are rather flat in $q^2$, which reduces the impact of
energy-resolution uncertainties. Alternatively, one may consider the following
two observables, closely related to $R_6$:
\begin{equation}
\label{eq:R6p}
\bar R_6[a,b] = \frac{\int_a^b A_{FB}^{\mu}}{\int_a^b A_{FB}^{e}}
    = R_6[a,b] / R_{K^*}[a,b] , \qquad \qquad
  R'_6 = \langle P_2^{(\mu)}\rangle/\langle P_2^{(e)} \rangle,
\end{equation}
where $A_{FB}$ is the forward-backward asymmetry and $P_2$ is the corresponding ``optimized observable'' defined in
ref.~\cite{Descotes-Genon:2013vna}.
For example, $\bar R_6$ has numerator and denominator normalised to the
respective integrated decay rate, such that a double ratio with
$\Gamma(J/\psi \mu \mu) / \Gamma(J/\psi e e)$ may be considered, similarly
to the $R_K$ and $R_{K^*}$ measurements.
$R_6$, $\bar R_6$, and $R_6'$  are theoretically equally clean
and have the same $C_{10}^\mu/C_{10}^e$
analyzing power as $R_6$.  The extreme insensitivity of $R_6^\prime$ to the values
of $C_9^\ell$ displayed in~\ref{tab:R6} and Fig.~\ref{fig:R6} is
remarkable and turns these observables into the optimal analyzer
of lepton nonuniversality in $C_{10}^\ell$.
However, the relatively larger sensitivity to $C_9^\ell$
in $\bar R_6$ relative to $R_6'$ is strongly correlated with that in $R_{K^*}$,
such that a joint measurement of $R_{K^*}$ and $\bar R_6$ can still
determine $C_{10}^\mu$ precisely.
By contrast, the observable $Q_2$ defined in \cite{Capdevila:2016ivx}
is sensitive to form factor ratios. It is a clean null test of lepton
flavour universality, but does not provide a similarly precision determination of
its violation, cf the Table.

\begin{figure}[h]
\begin{tabular}{cc}
  \includegraphics[width=90mm]{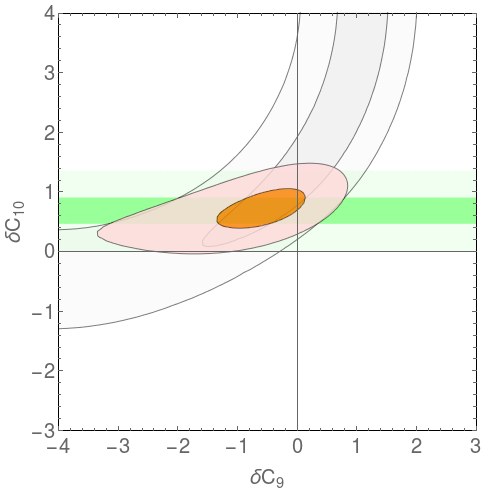}
\end{tabular}
\caption{Constraining power of the hypothetical measurement of $R_6^\prime$ in the bin $[0.045,~1.1]$ GeV$^2$ discussed in the main text,
represented at 1$\sigma$ and 3$\sigma$ by the horizontal green bands. This is shown overlaid to the 1$\sigma$ and 3$\sigma$ gray,
arc-shaped regions given by the fit to the LUV observables $R_K$ and $R_{K^*}$ discussed in Sec.~\ref{sec:fitRKRKs} and plotted in Fig.~\ref{fig:fitRKRKs}.
The combination of the two constraints results in the regions plotted with orange and light-red colors.  
\label{fig:parametric2}}
\end{figure}

In consequence, the choice among $R_6$, $\bar R_6$, and $R_6'$
should be guided by what minimizes experimental systematics. Most
excitingly, a useful determination of each of these should be possible with existing
data, both at LHCb and Belle. We illustrate this in Fig.~\ref{fig:parametric2} with the impact in the $(C_9^\mu,~C_{10}^\mu)$-plane of
a hypothetical measurement of $R_6^\prime=0.80(5)$ in the bin $[0.045,~1.1]$ GeV$^2$, shown together with the result of the fit to $R_K$ and $R_{K^*}$ and the combined result. 
We end by noting that we have not considered in our discussion of these new observables the effect of the electromagnetic radiative corrections, which can 
be more severe for the ultralow bin. However, as discussed in~\cite{Bordone:2016gaq} these can be minimized by choosing a more suitable
bin starting at 0.1 GeV$^2$, for which all the conclusions above will still apply.

\section{Conclusions}

In this paper we have analysed the deficits, with respect to the Standard-Model expectations, of the lepton-universality ratios $R_{K^*}$ recently measured by the LHCb. We first described the structure of the new-physics contributions to the $B\to K\ell\ell$ and $B\to K^*\ell\ell$ rates and the resulting correlations induced between $R_K$ and $R_{K^*}$. We then presented the predictions of $R_{K^*}$ in the bins of interest for the SM and in benchmark scenarios of new physics using a model-independent approach for the description of the $B\to K^*\ell\ell$ amplitude. The cleanness of these observables, with respect to hadronic uncertainties, is discussed and demonstrated numerically. Furthermore, we provide numerical formulas for the dependence of $R_{K^{(*)}}$ as functions of the Wilson coefficients which can be useful for phenomenological applications. Finally, we discarded the primed operators  $\mathcal O_{9(10)}^{\prime \ell}$ as they induce a correlation between $R_K$ and $R_{K^*}$ that is opposite to the observed one.   

Next, we perform fits to different sets of data within a frequentist approach in which the hadronic uncertainties are systematically included and minimized in the fits. We begin with the analysis of $R_{K^{(*)}}$ only, finding that the best fit is provided by combinations of the operators  $\mathcal O_{9(10)}^\ell=\bar s\gamma^\mu b_L~\ell\gamma_\mu(\gamma_5) \ell$ operators with a significance with respect to the Standard Model of $\sim4\sigma$. An important conclusion of this analysis is that a lepton-specific contribution to $\mathcal O_{10}$ is essential to understand the data. Under the hypothesis that new-physics couples selectively to the muons, we then present a series of fits to other $b\to s\mu\mu$ data with a conservative error assessment. First we include only the branching fraction of the $B_s\to\mu\mu$ decay, which provides a significant constraint that drifts the best fit point to a ``chiral'' left-handed solution $\mathcal O_{9}^\ell-\mathcal O_{10}^\ell$, but does not significantly increase the tension of the data with the SM. 

In the last fit we include all the angular observables measured by LHCb, ATLAS, CMS and Belle for the bins below 6 GeV$^2$. As previous fits in the literature, measurements of the angular observables tend to pull the Wilson coefficient $\delta C_9^\mu$ to negative values with a significance of more than 3$\sigma$, while rendering $\delta C_{10}^\mu$ consistent with 0 at $\sim1\sigma$. The results of these fits, rephrased in terms of $C_{L/R}^\mu=(C_9^\mu\mp C_{10}^\mu)/2$, show that $C_L^\mu$ is very well determined by the fit $\delta C_L^\mu\simeq-1$ while $C_R^\mu$ is not. We have also noted, however, the potential vulnerability of the results of this fit to underlying assumptions regarding the hadronic uncertainties, as it has been discussed extensively in the literature. In order to quantify this, we have made a study of the robustness of the tensions of the SM with the data for different assumptions regarding the size of the hadronic uncertainties. We emphasize again that this problem is absent in the fits to only $R_{K^{(*})}$ and $B_s\to \mu\mu$ and the respective statistical information inferred from the $\chi^2$ is driven, in this case, by experimental uncertainties only so that its interpretation is free from theoretical ambiguities.    

Therefore, we introduce and discuss a group of lepton-universality ratios all related to the coefficient of the angular distribution $I_6$ (or the forward-backward asymmetry) which are very sensitive to lepton nonuniversality in the Wilson coefficient $C_{10}^\ell$ and are largely insensitive to new-physics contributions to $C_9$ in the low-$q^2$ bin. A measurement of these observables would break the correlation between the lepton-specific effects in $\mathcal O_9^\ell$ and $\mathcal O_{10}^\ell$ observed in the fits to $R_K$ and $R_{K^*}$. In light of the fact that the two observables  $P_2$ have been measured by the LHCb for electrons and muons respectively at very low $q^2$ (though using a different binning), one is tempted to think that a measurement of the ratios proposed in this paper is feasible in the near future. We present prospects of the impact of such a measurement in the $(C_9^\ell,~C_{10}^\ell)$ plane, which could help clarifying the nature of the observed effects and contribute to the discovery of the new physics in $b\to s\ell\ell$ transitions.

\section{Acknowledgments}
We would like to thank Kostas Petridis and Luiz Vale Silva for helpful
discussions.
This work is partly supported by the National Natural Science Foundation of China under Grants  No. 11375024 and No. 11522539 and by the US Department of Energy under grant DE-SC0009919. SJ is supported by an IPPP Associateship,
STFC consolidated grant ST/L000504/1, and a Weizmann
Institute ``Weizmann-UK Making Connections'' grant, and would like to thank
STFC for past provision of HPC computing time.

\bibliography{RKs.bib}
\end{document}